\documentstyle[epsf,rotate,aps,multicol,prb]{revtex}

\include{psfig}

\begin{document}

%\draft

\title {\bf Structural properties and quasiparticle energies 
of cubic SrO, MgO and SrTiO$_3$}

%\vspace*{1cm}

\author{Giancarlo Cappellini}
\address{INFM Sezione di Cagliari and Dipartimento di Fisica,
Universit\`a di Cagliari, Cittadella Universitaria,
Strada Prov.le Monserrato-Sestu km.0.700,
I-09042 Monserrato (Ca), Italy}

\author
{Sophie Bouette-Russo, Bernard Amadon, Claudine Noguera and Fabio Finocchi}
\address{
Laboratoire de Physique des Solides, UMR CNRS 8502,
B\^atimemt 510, Universit\a'e Paris-Sud, 91405 Orsay, France }

\maketitle

\vspace*{2cm}

\begin{abstract}
\noindent
The structural properties and the band structures of the charge--transfer
insulating oxides SrO, MgO and SrTiO$_3$ are
computed both within density functional theory in the local density 
approximation (LDA) and in the Hedin's GW scheme 
for self-energy corrections, by using a model dielectric function, 
which approximately includes local field and dynamical effects. 
The deep valence states are shifted to higher binding energies 
by the GW method, in very good agreement with photoemission spectra. 
Since in all these oxides the direct gaps at high-symmetry points 
of the Brillouin zone may be strongly sensitive to the actual
value of the lattice parameter $a$, already at the LDA level,
self-energy corrections are computed both at the theoretical and
the experimental $a$.
For MgO and SrO, the values of the transition energies between
the valence and the conduction bands is improved
by GW corrections, while for SrTiO$_3$ they are overestimated.  
The results are discussed in relation to the importance of
local field effects and to the nature of the electronic states in these
insulating oxides.
\end{abstract}

%\begin{multicols}2

%\narrowtext

\newpage

\section{Introduction}

Aside from being the major constituent of the outer crust of
the earth, oxides are of 
considerable interest for their wide applications in various 
technological fields such as electrochemistry, catalysis or 
microelectronics.\cite{Noguera}. 
They present very diverse electronic ground states
which range from superconducting, metallic,
semi--conducting (intrinsic, charge density-- or spin density--like)
to insulating. 
>From a theoretical point of view, a proper description 
of their electronic properties is still an active research subject.
On the one hand, several transition metal oxides require an inclusion
of many-body effects beyond the available effective 
one-electron approaches, such as the density functional theory (DFT) 
or the {\it ab initio} Hartree-Fock (HF) method. 
On the other hand, even for insulating charge transfer
oxides, such as MgO and SrO, whose
electronic ground state is well accounted for in a one-electron picture, 
the calculation of quasiparticle (QP) energies within the DFT or the
HF method gives rather poor results. Indeed, one has to face the
so-called energy band gap problem \cite{Bechs92}. 
Indeed, when using the Kohn-Sham eigenvalues of the highest occupied 
and the lowest empty states, the resulting gap usually 
underestimates the experimental data. On the other hand, the corresponding 
difference between the HF eigenvalues overestimates the gap severely.
Alternative ways to calculate the band gap in infinite systems
have been found, in the framework of HF or DFT. 
One of them is based on total energy differences between ground and 
charge transfer states \cite{Noguera00}. The other one relies on the
value of the gap found in the one electron spectrum of the ionized or
hole-doped system \cite{Mackrodt96}. 
However, such methods are unable to provide the full QP spectrum.

In the last decade, due to the implementation of angular resolved
photoemission experiments able to cope with charging effects 
on insulating oxides, detailed QP
spectra of MgO\cite{Tjeng}, SrTiO$_3$\cite{Brookes} 
or TiO$_2$\cite{Raikar,Hardman} among other oxides, have become available. 
The experimental achievements demand
systematic improvements in the theoretical description 
of the electronic excitations of simple oxides.

In semiconducting and ionic crystalline compounds,
corrections to the DFT eigenvalues, as obtained from the Kohn-Sham 
equations, may be estimated by first-order many-body perturbation 
theory \cite{Hedin,Hybersten,Godby}, with respect to the difference 
between the exchange-correlation (XC) self-energy and the corresponding 
XC potential of the DFT, for which various approximations are available. 
Within the same approach, the self-energy operator is treated in 
the GW approximation proposed by Hedin\cite{Hedin}.
Such calculations have been carried out with success in several
semiconducting and insulating crystals
\cite{Hybersten,Godby,Rohlfing98,Shirley_3,Kralik,Schoenberg}, 
permitting the prediction of the quasiparticle energies and the 
calculation of optical properties of real solids.

One of the crucial points in such calculations is the evaluation of the 
self-energy operator, which is numerically very expensive, because of the 
required knowledge of the full inverse dielectric matrix of the system
under study.
This is a bottleneck for the application of the method to systems with a large 
number of atoms in the unit cell, such as surfaces \cite{Pulci98,Rohlfing99}
and defective or low--symmetry solids.
In order to reduce the remarkable numerical effort and be able
to predict electronic excited states in a wider range of real systems, 
one may use efficient GW methods, in which a model dielectric function 
is used to mimic the screening properties of the system under study. 

Pioneering work in this direction has been performed by Bechstedt 
and Del Sole in a tight-binding scheme \cite{Bechstedt88} and 
by Gygi and Baldereschi \cite{Gygi89} 
within the DFT in the Local Density Approximation (LDA)
for the exchange and correlation energy. 
The key of their work relies {\it(i)} on the use of a model dielectric 
function, which mimics the main screening properties of semiconductors, 
and {\it(ii)} on an approximated self-energy operator in which both
local fields and dynamical screening -- giving rise to
opposite contributions to band gaps--  are neglected.
\cite{Hybersten,Bechstedt88,Gygi89} For a large family of
semiconductors, the band gaps obtained through these simplified methods 
show good accordance both with experiments
and full GW results, in which the screening properties of
the system are calculated from first principles.
More recently, an efficient GW method that approximately 
includes dynamical-screening and local field effects, without increasing the 
computational effort, has been formulated \cite{Bechstedt92}
and proved to reproduce quasiparticle energies and band gaps
for either bulk zincblende solids (ranging from Si, GaAs, AlAs 
to the more ionic ZnSe)\cite{Palummo95} or non cubic 
systems (BN\cite{Cappellini96}, SiC polytypes\cite{Wenzien1,Wenzien2}) 
satisfactorily.  The method is based on a linear expansion
of the dynamical contribution to the exchange-correlation self-energy,
on the use of a model dielectric function, and on a local ansatz for the
treatment of the intrinsically non local screened interaction.
Depending on the system, the speedup of the calculation
may result of two order of magnitude with respect to 
corresponding full GW ones. Moreover, this kind of
GW calculations is feasible on a standard workstation.

The aim of the present paper is to study the ground state
properties of cubic magnesium oxide, strontium oxide and
strontium titanate, in the DFT-LDA scheme, as well as
their quasiparticle band structures within the efficient GW scheme.
On one hand, no previous GW calculation of SrO or SrTiO$_3$ exists,
to our knowledge. On the other hand,
these charge transfer oxides are characterized by different
fundamental gaps, ranging from $\simeq$ 3 eV (SrTiO$_3$) to
$\simeq$ 7.7 eV (MgO), as given by experimental data.
Their bandstructures may substantially differ 
from those of the semiconductors cited above, 
due to the presence of strongly localized electronic states.
Thus, such a study may also represent a test for the approximate GW scheme,
originally proposed for small-gap semiconductors,
to show its capabilities relatively to a different class of physical systems.

The paper is organized as follows. The theoretical framework
is depicted in Section II and it is followed, in Section III,
by a description of the computational ingredients used for
the description of structural and electronic properties of
MgO, SrO and SrTiO$_3$. The results obtained for these
systems are presented separately in Sections IV, V and VI,
respectively. Finally, in Section VII we present
our conclusions and the perspectives opened by our work.

\section {GW corrections}

The energy $E_{n\vec k}$ of a quasiparticle 
in a band $n$ at a given $\vec k$ point in the Brillouin zone
can be obtained through the equation: \cite{Hybersten,Fulde91,Godby}
\begin{equation}
\left[ -{\nabla^2 \over 2} + V_{\rm ext}({\vec r}) + V_{\rm H}
(\vec r) \right] \phi_{n \vec k}(\vec r ) \,  + 
\, \int d^3\vec r^\prime \, 
\Sigma(\vec r ,\vec{r^\prime}; 
E_{n \vec k}) \phi_{n \vec k}(\vec{r^\prime}) \, = \, 
E_{n \vec k} \phi_{n \vec k}(\vec r) 
\end{equation}
which is written in atomic units ($e$=$\hbar$ =$m_e$=1).
$V_{\rm ext}(\vec r)$ is the external potential originated by the ions, 
$V_{\rm H}(\vec r)$ is the Hartree potential and
$\Sigma$ is the mass operator, usually called the self--energy operator,
which is in general non local, non-Hermitian and energy dependent. 
In principle it contains the exchange and
correlation effects and is state (${n \vec k}$) dependent.
One can compare the above equation to that of Kohn and Sham \cite{KS65}:
\begin{equation}
\left[ -{\nabla^2 \over 2} + V_{\rm ext}(\vec r) + V_{\rm H}(\vec r)
+ V_{\rm xc}(\vec r) \right] \psi_{n \vec k}(\vec r)  =  
 \, E_{n \vec k}^{(0)} \psi_{n \vec k}(\vec r)   
\end{equation}
where $V_{\rm xc}$ is the exchange--correlation potential of the DFT
(for which several approximation are available),
and the superscript $^{(0)}$ will denote the quantities computed at
the DFT level from now on. 
$E_{n \vec k}^{(0)}$ is usually considered as a first approximation
to the quasiparticle energy.  
The failure of this scheme for evaluating transition energies
$E_{n^\prime \vec k} - E_{n \vec k}$ between different quasiparticle states
in solids has been evidenced by many authors \cite{Bechs92,Hybersten,Godby}. 

One can correct the DFT eigenvalues $E^{(0)}_{n \vec k}$,
in a first-order perturbation theory with respect
to $\Sigma_{n {\vec k}}-V_{\rm xc}$, and estimate the QP energies as:
\begin{equation}
E_{n\vec k} \, \simeq \, E_{n\vec k}^{(0)} + < \psi_{n\vec k} |
\Sigma(\vec r ,\vec r^\prime; E_{n\vec k}) - V_{\rm xc}(\vec r) 
| \psi_{n\vec k} >
\label{eq:3}
\end{equation}
The evaluation of $\Sigma(r,r^\prime; E_{n\vec k})$ is in general a 
very difficult task. A possible approximation is the GW one, \cite{Hedin}
in which a perturbation expansion for the self-energy 
can be constructed and stopped at the first order:
\begin{equation}
\Sigma(\vec r,\vec{r^\prime}; \omega) \, = \, i \int_{-\infty}^{+\infty}
{d\omega^\prime 
\over 2\pi} {\rm e}^{+i\delta \omega^\prime} \, G(\vec r,\vec{r^\prime}; 
\omega +\omega^\prime) \, W(\vec r,\vec{r^\prime}; \omega^\prime)
\label{eq:4}
\end{equation}
where $G$ is the one-particle Green function, $W$ is the screened
Coulomb interaction and $\delta=0^+$. 
Following Bechstedt and coworkers,\cite{Bechs92,Bechstedt92} 
the real part of the self-energy operator (which is relevant for the 
corrections to the QP bandstructure)
may be separated in a dynamic and a static contribution, and written as:
\begin{equation}
\Re \, \Sigma(\vec r ,\vec r^\prime;\omega) = 
\Sigma^{\rm sex}(\vec r ,\vec{r^\prime}) +
\Sigma^{\rm coh}(\vec r ,\vec{r^\prime}) + 
\Sigma^{\rm dyn}(\vec r ,\vec{r^\prime}; \omega)
\label{eq:5}
\end{equation}
$\Sigma^{\rm sex}$ and $\Sigma^{\rm coh}$ are the static 
screened--exchange (SEX) and the coulomb--hole (COH) terms, respectively, 
and $\Sigma^{\rm dyn}$ the energy dependent contribution.
Although the previous equation is purely formal, it is the
starting point of simplified GW schemes, since one can work out
approximated expressions for each contribution separately.
The most expensive step towards the evaluation of GW corrections
to $E_{n \vec k}^{(0)}$ is the calculation of the energy-dependent screened
interaction $W(r, r^\prime; \omega)$. Although many improvements
have been proposed for an efficient calculation of the dielectric
response function from first principles, \cite{Hybersten88,Reining1} 
still this stage is
both computer time and memory very demanding when a large basis set
is used for the description of the electronic structure.
In particular, this is the case of oxides in a plane-wave
pseudopotential approach.
On the other hand,
the improvement of DFT--LDA band structures obtained by using
schemes originally formulated on phenomenological basis, such as the 
scissor operator or the Slater's $\alpha$ potential, \cite{Bechs92}
showed that the essential physics can be described through the use
of effective or state-dependent potentials. From a more 
fundamental point of view, one can work out a model 
for the screening properties of the system, and look to what extent
the computed QP spectra account for the observed
experimental values. Possibly, the model should also be simple enough  
to be generalized to a large number of different systems.
Following previous studies, \cite{Bechstedt92,Wenzien1}
we use a model dielectric function to estimate both
the static and the dynamical contributions to the self-energy 
(Eq.\ref{eq:5}), since it is the key ingredient entering
the efficient GW approximations.

\subsection{Static Screening and Local Field Effects}

For the diagonal part of the screening function, we use
an analytical model originally proposed by Bechstedt and Del Sole: 
\cite{Bechstedt88}
\begin{equation}
\epsilon (q,\bar \rho )= 1+\left[ {1\over \epsilon_{\infty}-1}
+ {q^2 \over q_{TF}^2} + 
{3 q^4 \over 4 k_F^2 \, q_{TF}^2}\right] ^{-1} \label{eq:6}
\end{equation}
where the Fermi vector $k_F$ and the Thomas-Fermi wave-vector
$q_{TF}$ depend on the average electron density $\bar\rho$. 
This expression mimics the free electron gas behavior
at high $q$. At $q=0$, Eq.\ref{eq:6} gives the value of the optical
dielectric constant $\epsilon_{\infty}$ of the oxide.
Moreover we use a local {\it ansatz} for the static Coulomb screened 
interaction, analogously to that of Hybertsen and Louie \cite{Hybersten88}:
\begin{equation}
W(\vec r,\vec{r'}) = {1\over 2}
[W^{h}(\vec r-\vec r^\prime ;\rho(\vec r))+ 
W^{h}(\vec r-\vec r^\prime ;\rho(\vec r^\prime ))] 
\label{eq:7}
\end{equation}
where $W^{h}$ is the screened Coulomb interaction of a virtual
homogeneous system characterized by a finite optical dielectric constant.
\cite{Bechstedt88,LevLeo} Quasiparticle energies
determined from the above {\it ansatz} result in good accordance with
full GW calculated ones.\cite{Bechstedt92,Hybersten88}
It permits, together with Eq.(6), to obtain a simple
analytical expression for the static part of
the Coulomb hole self energy $\Sigma^{\rm coh}$. 
The explicit formula has been given elsewhere. \cite{Cappellini93}
For the calculation of $\Sigma^{\rm sex}$
in Eq.\ref{eq:5}, the Fourier transform of $W(\vec r,\vec r^{\,\prime})$ 
is needed, namely $\tilde W(\vec k + \vec G, \vec k + \vec G^\prime)$. 
In order to reduce sharply the time needed to compute this term,
we proceed in two steps. Firstly, we retain only 
the diagonal terms in the screened coulomb interaction. Secondly,
we account for local field  effects ({\it i.e.} non diagonal terms with
$\vec G \ne \vec G^\prime$ in $\tilde W$)
by using suitable state-dependent densities to compute $k_F$ and $q_{TF}$.
Given a state $n\vec k$,  we define the effective electron density
$\tilde \rho_{n\vec k}$, which determines the typical screening lengths 
$q_{TF}^{-1}$ and $k_F^{-1}$, relevant for the estimation of
$E_{n,\vec k}$ (Eq.\ref{eq:3}) as the expectation value of the ground state
electronic density $\rho(\vec r)$ on the actual state:
\begin{equation}
\tilde\rho_{n\vec k}=\int d^3 {\vec r} \, \rho (\vec r)  \,
\vert \psi_{n\vec k}(\vec r) \vert^2 \, . 
\label{eq:8}
\end{equation}
Since the relevant screening lengths $k_F^{-1}$ and $q_{TF}^{-1}$ are
increasing functions of $\tilde\rho_{n\vec k}$, the latter is intended
to give an approximate description of the local fields in $\Sigma^{\rm
sex}$ through an enhanced screening for the electronic states 
which contribute effectively to the total electron density.
In the oxides considered here, the effective densities 
$\tilde\rho_{n\vec k}$ range from few tenths of the 
mean valence electron density $\bar\rho=Z_{\rm val}/V$ 
for conduction states up to 4--5 $\bar\rho$ for deep valence states.
As a consequence, the lengths $k_F^{-1}$ and $q_{TF}^{-1}$ 
are shorter for the screening of valence band states
than for conduction band states. 
This difference is enhanced with respect to more covalent semiconductors, 
such as Si and GaAs \cite{Bechstedt92}, in which the
valence electron distribution is more delocalized and less dissimilar
from that of the conduction states than in oxides such as MgO.

This approximate scheme for treating local field
effects was analyzed in detail and its reliability
was shown through a comparison with full GW calculations,
in the case of both small-- and wide--gap semiconductors
\cite{Bechstedt92,Palummo95}.
In Ref.\onlinecite{Palummo95}, it was also shown that 
the use of a non-diagonal model dielectric function does not clearly 
improve band gaps while bringing additional computational costs.

\subsection{Dynamical contributions}

The diagonal matrix element of the dynamical part of the
self energy (the third operator on the
right side in Eq.\ref{eq:5}) on the state $\psi_{n \vec k}$:
\begin{equation}
\Sigma^{\rm dyn}_{n\vec k}(\omega) = \langle \psi_{n\vec k} \vert 
\Sigma^{\rm dyn}(\vec r, \vec{r^\prime} ;\omega)
\vert \psi_{n\vec k}\rangle \nonumber
\end{equation} 
can be linearly expanded in energy around the DFT value 
 $E^{(0)}_{n\vec k}$ , with linear coefficient:
\begin{equation}
\beta_{n\vec k} = \langle \psi_{n \vec k} \vert 
\left( {\partial\Sigma \over \partial\omega }\right)_{E^{(0)}_{n \vec k}} 
\vert \psi_{n \vec k} \rangle.
\nonumber
\end{equation}
This approximation for the energy dependence of the self-energy
operator turns out to be valid in a wide energy range, as
confirmed by the results of full GW calculations on bulk Si.\cite{Hybersten}
Solving the Dyson equation within the first-order perturbation theory
with respect to $\Sigma - V_{\rm xc}$, 
and using Eqs.\ref{eq:3} and \ref{eq:5}, the QP shifts for the
$\psi_{n\vec k}$ state hence read:

\begin{equation}
E_{n\vec k} - E^{(0)}_{n\vec k} =  
{ \Sigma^{\rm coh}_{n \vec k}+\Sigma^{\rm sex}_{n \vec k} + 
\Sigma^{\rm dyn}_{n \vec k}(E^{(0)}_{n\vec k}) \, - \,
\langle \psi_{n \vec k} | V_{\rm xc} | \psi_{n \vec k} \rangle 
\over  1+\beta_{n\vec k} },
\label{eq:13a}
\end{equation}

\noindent
where $\Sigma^{\rm coh}_{n \vec k}$ and $\Sigma^{\rm sex}_{n \vec k}$ 
are the expectation values of the static COH and SEX contributions 
to the self energy, respectively.
>From the GW expression of $\Sigma$, replacing G by G$^{(0)}$,
computed at the DFT level \cite{Bechstedt92} :

\begin{equation}
\Sigma^{\rm dyn}_{n\vec k}(E^{(0)}_{n\vec k})  = \sum_{n'\vec k'}
\int {d^3 {\vec q} \over (2\pi)^{3}} \, 
\vert B_{nn'}^{\vec k\vec k'}(\vec q) \vert^2
{4\pi \over q^2}(E^{(0)}_{n\vec k}-E^{(0)}_{n'\vec k'}) 
P\int_0^{\infty}{d\omega\over \omega\pi}
{ \Im \big( \epsilon^{-1} (\vec q, \tilde \rho_{n\vec k} ,\omega ) \big) \over 
\omega - {\rm sgn}(E^{(0)}_{n'\vec k'}-\mu)(E^{(0)}_{n\vec k}
-E^{(0)}_{n'\vec k'}) }
\label{eq:12}
\end{equation}

\begin{equation}
\beta_{n\vec k} = \sum_{n'\vec k'}
\int {d^3 {\vec q} \over (2\pi)^{3}} \, 
\vert B_{nn'}^{\vec k\vec k'}(\vec q) \vert^2 {4\pi \over q^2}
P\int_0^{\infty}{d\omega\over \pi}
{ \Im \big( \epsilon^{-1} (\vec q,\tilde \rho_{n\vec k},\omega ) \big) \over 
\left[ \omega - {\rm sgn}(E^{(0)}_{n'\vec k'}-\mu)(E^{(0)}_{n\vec k}
-E^{(0)}_{n'\vec k'})\right]^2 }
\label{eq:13}
\end{equation}

where
\begin{equation}
B_{nn^\prime }^{\vec k \vec{k^\prime} }(\vec q) \, = \,
\int d^3 {\vec r} \, \psi_{n\vec k}^\ast (\vec r) \, 
{\rm e}^{ i{\vec q} \cdot {\vec r}} \, \psi_{n^\prime \vec{k^\prime} } (\vec r)
\label{eq:b}
\end{equation}

The model $\epsilon^{-1}$ is diagonal in $q$ space
and the local field effects are
accounted for through the use of the state averaged density 
$\tilde\rho_{n\vec k}$ (Eq.\ref{eq:8}). 
The $\omega$-dependent dielectric
function is obtained from the static one by means of
the plasmon-pole approximation (PPA)\cite{Hybersten}.
The PPA has been widely used to compute semiconductor and 
insulator quasiparticle bandstructures, even in the energy range
of semicore states \cite{Rohlfing98,Shirley_3,Kralik}.
In order to account for the differences
$E^{(0)}_{n\vec k}- E^{(0)}_{n'\vec k'}$
in Eqs.\ref{eq:12} and \ref{eq:13}, replacing them
with the state independent form 
$E^{(0)}_{n\vec k}- E^{(0)}_{n'\vec k'} \approx -q^2/2$
as in Ref.\onlinecite{Bechstedt92}, and
using the sum rule $\sum_{n^\prime \vec k^\prime}
\vert B_{n n^\prime}^{\vec k \vec k^{\,\prime}} (\vec q) \vert^2 = 1$, 
one obtains:

\begin{eqnarray}
\Sigma^{\rm dyn}_{n\vec k}(E^{(0)}_{n\vec k}) & \simeq & 
\int {d^3 {\vec q} \over (2\pi)^3 } \, 
P\int_0^{\infty}{d\omega\over \omega\pi}
{ \Im \big( \epsilon^{-1} (\vec q,{\tilde \rho_{n\vec k}},\omega )
 \big) \over 
\omega +{q^2\over 2}}
\label{eq:15}
\end{eqnarray}

\begin{equation}
\beta_{n\vec k} \simeq \int {d^3 {\vec q} \over (2\pi)^3 } \, 
{4\pi e^2\over q^2} P \int_0^{\infty}{d\omega\over \pi}
{ \Im \big( \epsilon^{-1} (\vec q, {\tilde \rho_{n\vec k}},\omega )
 \big) \over ( \omega +{q^2\over 2})^2 }
\label{eq:16}
\end{equation}

Therefore, once the spectrum and eigenfunctions obtained in the
framework of the DFT are known, 
the calculation of $\Sigma^{dyn}_{n\vec k}(E^{(0)}_{n\vec k})$
and $\beta_{n\vec k}$ according to  
Eqs.\ref{eq:15} and \ref{eq:16} needs no additional external 
parameters apart from the model dielectric function,
and results a simple task 
that can be carried out on a standard workstation.

We have made extensive checks to test the reliability 
of the approximations used to obtain Eqs.\ref{eq:15} and \ref{eq:16}, 
in the case of Silicon. Some transitions between 
high--symmetry points of some valence and conduction band states in Si 
are compared in Tables I to those obtained through full GW.
The energy of the transitions between high--symmetry points, 
as well as the fundamental gap in Si, are in very close agreement 
with those obtained from full GW calculations and with experiments 
for both methods.
As far as the QP eigenvalues in Si are concerned, the top of the
valence band $\Gamma_{25^\prime,v}$ is shifted downwards with respect to
the DFT values  of 0.28 eV. This can be compared
with the upward shift of 0.07 eV, as found by Godby {\it et al.}\cite{Godby}.

\begin{table}[p]
\begin{tabular}{|l|c|c|c|c|}
\noindent
 Si & full GW $^{(a)}$ & full GW $^{(b)}$ & this work &  Exp. \\
\hline\hline
$\Gamma_{1,v} - \Gamma_{25^\prime,v}$   & 12.04 &      & 12.09 & 12.5$\pm$0.6\\
$\Gamma_{25^\prime,v}-\Gamma_{15,c}$     &  3.35 & 3.30  & 3.33  &    3.4 \\ 
$\Gamma_{25^\prime,v}-\Gamma_{2^\prime,c}$&  4.08& 4.27  & 4.25  &    4.2 \\ 
\hline
$L_{2^\prime,v}-\Gamma_{25^\prime,v}$   &  9.79 &  & 9.98 &  9.3$\pm$0.4\\
$L_{1,v}-\Gamma_{25^\prime,v}$          &  7.18 &  & 7.27 &  6.7$\pm$0.2\\
$L_{3^\prime,v}-\Gamma_{25^\prime,v}$   &  1.27 &  & 1.25 &  1.2$\pm$0.2, 1.5\\
\hline
$\Gamma_{25^\prime,v}-L_{1,c}$   &  2.27 & 2.30  & 2.22 & 2.4$\pm$0.2, 2.1\\
$\Gamma_{25^\prime,v}-L_{3,c}$   &  4.24 & 4.11  & 4.15 &   4.15$\pm$0.1\\
\hline
$L_{3^\prime,v}-L_{1,c}$            &  3.54 &       & 3.47 &   3.45 \\
$L_{3^\prime,v}-L_{3,c}$            &  5.51 &       & 5.41 &   5.50 \\
\hline
$X_{4,v}-\Gamma_{25^\prime,v}$    &  2.99 &       & 2.98 &   3.3$\pm$0.2, 2.9\\
$\Gamma_{25^\prime,v}-X_{1,c}$     &  1.44 &       & 1.54 &   1.3 \\
\hline
$v-c$ minimal gap                    &  1.29 & 1.24  & 1.21 &    1.17 \\
\end{tabular}
\caption{ Computed vertical transitions and minimal valence--conduction
band gap for Silicon (in eV). $(a)$: Ref.\protect\onlinecite{Hybersten88}. 
$(b)$: Ref.\protect\onlinecite{Godby}.
The experimental values are quoted in Ref.\protect\onlinecite{Hybersten88}.  }
\label{tab:0a}
\end{table}

\section{Computational details}

The Density Functional calculations are carried out within the 
LDA for exchange and correlation\cite{Ceperley81}.
The Kohn--Sham orbitals are expanded in a plane--wave
basis set. Special care has been used in constructing the
pseudopotentials for cations (Sr, Ti, Mg), in order to avoid the occurrence of
ghost states \cite{Gonze} and to assure an optimal transferability
over a wide energy range. 
We find that the inclusion of semicore states, such as $4s$ and $4p$
states for Sr, and $3s$ and $3p$ states for Ti, greatly improves the
transferability of the pseudopotentials, and is at the same time necessary 
to take into account the hybridization of cation semicore states with
O $2s$ states. 

Angular components up to $l=2$ are included. 
The scheme of Martins and Troullier is used \cite{Troullier} to generate
separable soft norm--conserving pseudopotentials, with core radii
(bohr): 2.00 (Sr, $4s$), 1.50 (Sr, $4p$), 1.90 (Sr, $4d$); 
1.30 (Ti, $3s$), 1.40 (Ti, $3p$), 1.80 (Ti, $3d$); 
1.38 (O, $2s$), 1.60 (O, $2p$), 1.38 (O, $3d$);
2.00 (Mg, $3s$, $3p$, $3d$). The cutoff energy needed to obtain 
a convergence better than 0.1 eV of both total energy 
and Kohn-Sham eigenvalues is found to
be equal to 50 Ry for MgO, 60 Ry for SrO and 80 Ry for SrTiO$_3$.

The use of 10 special $\vec k$ points \cite{Monkhorst} 
for charge integration in the
irreducible wedge of the Brillouin Zone (IBZ) is sufficient to
achieve a good accuracy for the computed total energy;
for instance, the total energy changes by 
less than 10 meV and the fundamental band gap by less than 0.1 eV
when using 10 instead of 6 special points to sample the IBZ.

The main bottleneck in the ab--initio GW calculations is the
calculation and the inversion of the full dielectric matrix 
from the eigenfunctions and eigenvalues of the system. 
Typically, this task takes approximatively $75\%$ of CPU time needed 
for the calculation of the self--energy correction for a single state
\cite{Bechs92}. As a result of the use of a model screening,
the computational cost and memory requirement are strongly reduced,
and are comparable to those of calculations carried out within the DFT. 
Efficient GW 
calculations can thus be carried out also on a common workstation.
\cite{note_dec}

The evaluation of the GW corrections to the LDA bandstructures, 
and especially of the SEX contribution $\Sigma^{\rm sex}$ to
the self--energy, needs some care, due to the
presence of an integrable divergence. A reduction of the
numerical effort by using a limited number of $k$ points in the IBZ
is possible through the method proposed by Gygi and Baldereschi
\cite{Gygi86}. In our calculations, the regularization is
performed, for all the $\vec k$--points for which $\Sigma^{\rm sex}$ is 
calculated, by
transforming the integral over the BZ of the diverging contribution
into an integral of a function $F(\vec k)$ (periodic over the BZ),
that can be computed analytically,
plus a discrete sum over special $\vec k$-points 
of a smooth function. For the sake of conciseness, in the
case of the bare exchange term (the generalization to the case of 
the diagonal screened exchange in cubic systems is straigthforward), 
the transformation reads:

\begin{equation}
 \Omega \int_{BZ} {d^3k^\prime\over (2\pi )^3} \, 
{ \vert B_{nn^\prime}^{\vec k \vec k^\prime}(\vec g)\vert^2 \over 
  \vert \vec k^\prime \! -\!\vec k-\! \vec g \vert^2               } 
\, \simeq \, \Omega\int_{BZ} {d^3k^\prime\over (2\pi)^3} \,
F(\vec k^\prime) \, \vert B_{nn^\prime}^{\vec k\vec k^\prime}(0) \vert^2 
\; + \sum_{\vec k^\prime} \, w_{\vec k^\prime}  \left[ 
{\vert B_{nn^\prime}^{\vec k\vec k^\prime}(\vec g) \vert^2 \over 
 \vert \vec k^\prime\!  -\! \vec k\! -\! \vec g \vert^2 } -
\theta^{\rm BZ}(\vec k^\prime\! -\! \vec k \! -\!  \vec g) \, 
\vert B_{nn^\prime}^{\vec k\vec k^\prime}(0) \vert^2 
\, F(\vec k^\prime\! -\!\vec k)  \right]  .
\end{equation}

$\theta^{\rm BZ}(\vec q)=1$ if $\vec q \in {\rm BZ}$ and zero otherwise, 
$B_{nn^\prime}^{\vec k\vec k^\prime}(\vec g)$ has
previously been defined (Eq.\ref{eq:b}), and $w_{\vec k}$ is the weight
associated to the special point $\vec k$ in the IBZ.
By following this method, 10 special $\vec k$ points in
the IBZ are sufficient to get converged $\Sigma^{\rm sex}$
within few tens of meV.

Because of the presence of the static screened exchange term, 
each GW correction to a given $E^{(0)}_{n \vec k^\prime}$
scales as $N_{\vec k} N N_{\rm PW}$, where $N_{\vec k}$, $N$ and $N_{\rm
PW}$ are the number of ${\vec k}$ points used in the summation of the 
charge, of occupied electronic states, and of the plane waves 
used in the expansion of the Kohn--Sham orbitals, respectively.

\section{{\bf Magnesium oxide}}

%\subsection{Results}

The ground state properties obtained for the $O^5_h$ phase (rocksalt) of
magnesium oxide are summarized in Tab.\ref{tab:1a}. They are computed by
fitting the curve of total energy versus lattice parameter to the 
equation of state proposed by Murnaghan. One can note that our results 
compare well to other all-electron LDA calculations
\cite{Schoenberg,Taurian} and show slight discrepancies with respect
to experimental data: the computed lattice parameter is underestimated
by about 2\% and the cohesive energy is about 15 \% too large. 
We attribute these errors mainly to the use of the LDA.

\begin{figure}[p]
%\special{psfile=MgO_bands.ps 
%         hoffset=-20 voffset=20 hscale=40 vscale=40 angle=270}
% \vspace*{8.0cm}
\includegraphics{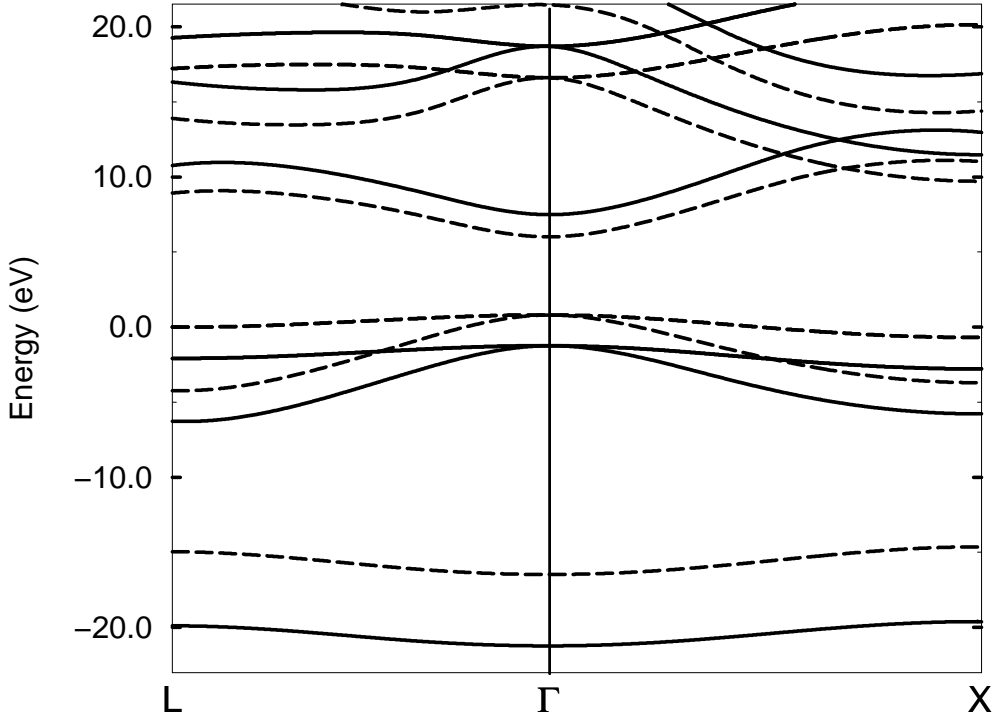}
%         hoffset=-20 voffset=20 hscale=60 vscale=60 angle=270}
\vspace*{10.5cm}
\caption{ Computed bands of cubic MgO at the theoretical lattice parameter. 
Solid lines: GW bands.  Dashed lines: LDA bands. The top of the
LDA valence bands is arbitrarily set to zero.}
\label{fig:mgo}
\end{figure}

The bandstructures, both in the LDA and including quasiparticle corrections,
are shown in Fig.\ref{fig:mgo} along high-symmetry lines in the BZ. 
In the following, we refer to the energy levels relative to 
the top of the valence band in the LDA calculation,
which is arbitrarily set to zero.
The bands are calculated at the theoretical equilibrium lattice parameter
$a_{\rm th}$=4.125 {\AA}. In Table \ref{tab:1b} we detail the 
energy differences between electronic states at high--symmetry points
in the BZ, computed either at the theoretical 
equilibrium lattice parameter $a_{\rm th}$=4.125 {\AA}
or at the experimental one $a_{\rm exp}$=4.211 {\AA}.
One can easily see that some of the direct transitions,
and in particular those at $\Gamma$, strongly depend upon the value 
of the lattice parameter.
This is true at the LDA level, while the GW
corrections are less sensitive to the actual value of $a$.
In ionic rocksalt compounds, the value of the fundamental gap 
at the $\Gamma$ point is indeed driven by the strong Madelung potential, 
which varies as the inverse of the lattice parameter.

In agreement with a previous full GW calculation \cite{Schoenberg}, 
the quasiparticle corrections consist
of a rigid shift of the conduction band (CB) with respect to the valence
band (VB) --- that is, application of a scissor operator, which is modulated by smooth
$\vec k$ dependent terms of the order of 10 \% of the rigid shift itself.
The total valence bandwidth $\Gamma_{15,v}-\Gamma_{1,v}$
changes from 17.3 eV in LDA to 20.1 eV with quasiparticle corrections.
Similarly, the $L^\prime$ and $X_1$ points move $\simeq$ 2 eV downwards
from the the valence band maximum (VBM), as one can see from Fig.\ref{fig:mgo}.
This improves the accordance with the experimental XPS spectra 
which show a main peak at 18 eV and a shoulder at 21 eV 
below the VBM \cite{Kowal77}.
The fundamental band gap passes from 5.21 eV to 8.88 eV.

In all these calculations, 
we assume the experimental value $\epsilon_\infty = 2.95$ \cite{Whited} 
for the static dielectric constant. Recently, through 
an {\it ab initio} calculation of the dielectric function within the
LDA, Shirley found $\epsilon_\infty = 3.03$, at the experimental
lattice parameter \cite{Shirley}. In order to test the sensitivity 
of our theoretical scheme to the choice of $\epsilon_\infty$,
which enters as a parameter in our model screening function 
defined in Eq.\ref{eq:6},
we performed an additional GW calculation by adopting $\epsilon_\infty =
3.50$. Consistently with a stronger screening, the direct transition 
energies at high--symmetry points are lowered by $\simeq 0.5$ eV at most.
 
It is interesting to note that the main corrections to the Kohn-Sham
eigenvalues of VB states come from the static screened exchange 
term $\Sigma^{\rm sex}$ (Eq.\ref{eq:8}), while the coulomb hole
contribution $\Sigma^{\rm coh}$ dominates the quasiparticle corrections
to the eigenvalues of the CB. 
For instance, $\Sigma^{\rm sex}_{\Gamma_1} \simeq -16$eV and 
$\Sigma^{\rm coh}_{\Gamma_1} \simeq -10$ eV at the bottom of the VB, 
while $\Sigma^{\rm sex}_{\Gamma_1} \simeq -5$ eV and 
$\Sigma^{\rm coh}_{\Gamma_1} \simeq -8$ eV at the bottom of the CB.
This is consistent with the 
more localized nature of the VB states with respect to 
the CB states. The same remark applies to SrO and SrTiO$_3$, which are 
discussed in the following sections.

As far as the interpretation of the main optical
transitions is concerned, we indicate in Table \ref{tab:1b} the
experimental values with their tentative assignement,
according to Sch\"omberger and Aryasetiawan \cite{Schoenberg}.
We agree with their interpretation, since our GW values, computed at the 
experimental lattice constant, differ by $\simeq 1$eV at most from theirs. 
Our calculations slightly overestimate the experimental 
transition energies.

\begin{table}[p]
\begin{tabular}{|l|l|l|l|}
\noindent
 MgO  & $a$({\AA}) & $B$(Mbar) & $E_{\rm coh}$ (eV) \\
\hline
{\rm Experiment}             & 4.211 & 1.55 \cite{Sangster} & 10.33 \\
\hline
{\rm Ref.\onlinecite{ChangCohen}}  & 4.191 & 1.46 &       \\
{\rm Ref.\onlinecite{Schoenberg}}  & 4.16  &      &       \\
{\rm Ref.\onlinecite{Taurian}}     & 4.09  & 1.71 & 10.67 \\
\hline
{\rm Present work}           & 4.125 & 1.56 & 11.80   \\
\end{tabular}
\caption{Structural properties of Cubic MgO. 
A comparison is made with other LDA calculations,
in a pseudopotential approach \cite{ChangCohen}
or using LMTO \cite{Schoenberg,Taurian} }
\label{tab:1a}
\end{table}
\begin{table}
\begin{tabular}{|l|l|l|l|l|}
\noindent
%{\rm cubic MgO}  & $\Gamma_1-\Gamma_{15}$ & $X_3-X_5^\prime$ 
% & $L^\prime_2-L_3$ & Minimal Gap \\
 MgO  & $E_{\rm g}(\Gamma)$ & $E_{\rm g}(X)$ 
 & $E_{\rm g}(L)$ & Minimal Gap \\
\hline
\rm Experiment\cite{Roessler}  & 7.7 & 13.3 & 10.8 & 7.7 ($\Gamma-\Gamma$)\\
\hline
HF+pol. \cite{Pandey91} & 8.21 & 15.79 & 11.48 & 8.21 ($\Gamma-\Gamma$)\\
LDA \cite{Schoenberg}  & 5.2  & 10.5  & 8.9  & 5.2 ($\Gamma-\Gamma$) \\
LDA+GW \cite{Schoenberg} & 7.7  & 13 & 11.4  & 7.7 ($\Gamma-\Gamma$) \\
LDA \cite{Shirley}   & 4.73  &  &   & 4.73 ($\Gamma-\Gamma$) \\
LDA+GW \cite{Shirley}& 7.81  &  &   & 7.81 ($\Gamma-\Gamma$) \\
\hline
this work            &     &      &     &                       \\
LDA$^\dag$           & 5.21 & 10.40 &  8.93 & 5.21 ($\Gamma-\Gamma$) \\
LDA$^\ddag$          & 4.61 & 10.26 &  8.46 & 4.61 ($\Gamma-\Gamma$) \\
LDA+GW$^\dag$        & 8.88 & 14.43 & 12.99 & 8.88 ($\Gamma-\Gamma$) \\
LDA+GW$^\ddag$       & 8.20 & 14.22 & 12.46 & 8.20 ($\Gamma-\Gamma$) \\
\end{tabular}
\caption{ Direct gaps of Cubic MgO (in eV). Our results are given both 
in the LDA and with quasiparticle corrections (LDA+GW), either at
the theoretical equilibrium lattice constant ($^\dag$) or at the
experimental one ($^\ddag$). The theoretical calculations
of Refs.\protect\onlinecite{Pandey91,Schoenberg,Shirley} are all carried out
at the experimental lattice constant. }
\label{tab:1b}
\end{table}

\section{{\bf Strontium Oxide}}

%\subsection{Results}

\begin{figure}[h]
% \special{psfile=SrO_bands.ps hoffset=-35 hscale=45 vscale=45 angle=270}
% \vspace*{-1cm}
 \includegraphics{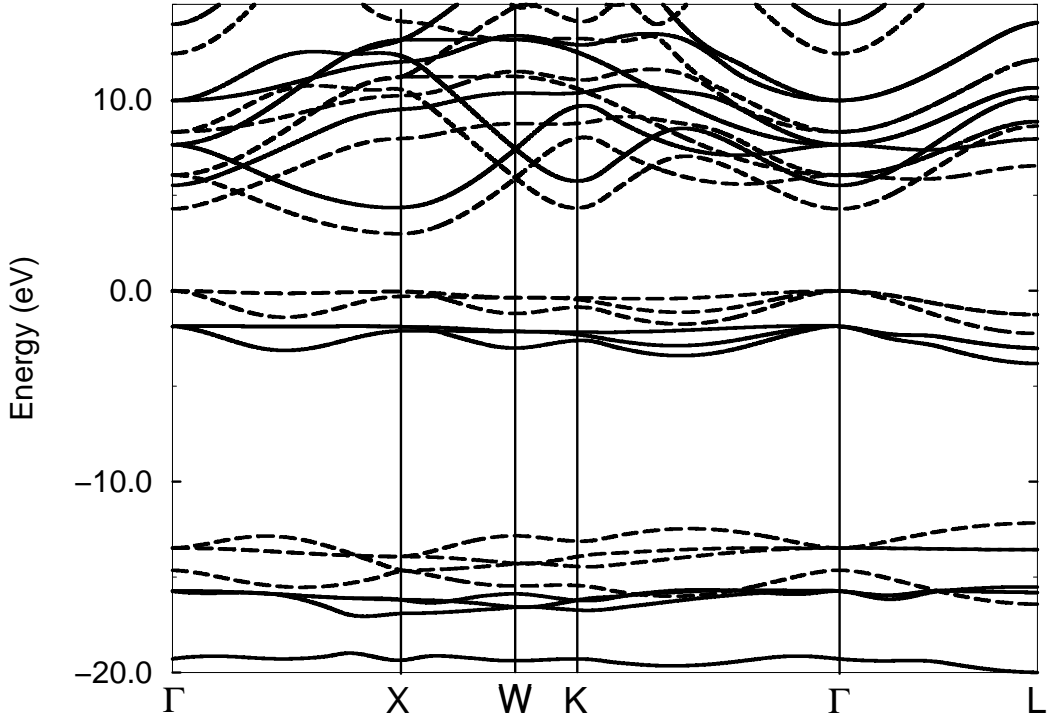}
 \vspace*{10.5cm}
 \caption{ Computed bands of cubic SrO, at the theoretical lattice
parameter. Dashed line: LDA bands.  Full line: GW bands. The zero
is arbitrarily fixed at the top of the valence LDA band.}
 \label{fig:sro1}
 \end{figure}

The ground state properties computed for the $O_h^5$ phase of SrO are
summarized in Tab.\ref{tab:2a}.
The bandstructures, computed in LDA and with QP corrections,
are shown in Fig.\ref{fig:sro1} along high symmetry directions. 

A deep band, originating mainly from Sr $4s$ states, is not shown in the
figure. It is found around -30.5 eV in the LDA calculation, with a
dispersion less than 1eV. The QP corrections shift this band
to lower energies, about 33 eV below the valence band maximum (VBM).

The bands of lowest energy shown in Fig.\ref{fig:sro1}
originate from Sr $4p$ and O $2s$ states. 
In the LDA, the bandwidth is about 4.3 eV, while the GW corrections 
push that band towards higher binding energies, and split it in two less
dispersive structures, of which the lower originates from O $2s$ 
states mainly, and the upper from Sr $4p$ states. 
The splitting between O $2s$ and Sr $4p$ levels
is essentially due to the larger $\Sigma^{\rm sex}$ contribution for
the former states, as a consequence of their stronger localization.

As far as the deep VB levels are concerned, XPS experiments
\cite{Kowal77} found two broad peaks at $\simeq -35$ eV ({\it A}) and 
$\simeq$ -17.5 eV ({\it B}) below the top of the valence band. 
The authors assigned the {\it A} structure to the band
arising from Sr $4s$ states, and the {\it B} structure 
to those coming from O $2s$ and Sr $4p$ states. 
The resolution, however, was too poor for them
to resolve the two latter contributions to peak {\it B}. 
The positions of the deep VB levels are badly
accounted for in the LDA, which would yield the {\it A} peak 
at -30.5 eV and a broader {\it B} structure around -13 eV.
On the other hand, Hartree--Fock calculations including correlation 
\cite{Pandey91} give a separation of about 6.5 eV between the O $2s$ 
and the Sr $4p$ states, which is likely overestimated.  
A better agreement with experimental data is obtained through the
inclusion of GW corrections, which shift the Sr $4s$ band 
to $\simeq$ -33 eV and give two contributions to the electronic
density of states at $\simeq$ -18 eV and $\simeq$ -14 eV with
respect to the VBM, about 0.5 eV wide. 
The latters seem consistent with the broad $B$
structure seen in XPS \cite{Kowal77}. 
A later XPS investigation \cite{vanDov80} found a splitting $\Delta_{AB}$
between the $A$ and the $B$ structure equal to 17.9 eV, which was interpreted 
as the difference in the binding energy of Sr $4s$ and $4p$ levels in SrO.
Since it seems difficult to disentangle the Sr $4p$ and the O $2s$
contribution to the $B$ structure, we compare the minimum and the maximum
splitting $\Delta_{AB}$ as obtained in our calculations.
In the LDA, $\Delta_{AB}$ ranges from 14.5 eV to 17.5 eV, while that obtained
by including the GW corrections ranges from 16 eV to 19 eV and better
agrees with the XPS data.

The upper part of the valence band results very similar in the LDA and
GW calculations. Its width is 2.2 eV in LDA and 2.1 eV after GW
corrections. The dispersions of both QP and LDA
eigenvalues along high--symmetry directions are very similar, too. 

The lower part of the conduction band drawn in the figures is shifted
upward by GW corrections, opening the LDA gap by about 3 eV, with
additional and small corrections depending on the particular band and
$\vec k$ point as one can see in Table (\ref{tab:2b}). In particular, the direct 
gap at $\Gamma$ ($X$) passes from $E_{\rm g}(\Gamma) = 4.29$ eV 
($E_{\rm g}(X) = 3.03$ eV) in LDA to $E_{\rm g}(\Gamma) = 7.54$ eV 
($E_{\rm g}(X) = 6.39$ eV) after GW corrections.  
It is important to note that 
the direct gap $\Gamma_{15}$ -- $\Gamma_{1}$ is strongly sensitive to
the precise value of the lattice parameter $a_0$: while the above value
$E_{\rm g}(\Gamma) = 4.29$ eV is calculated at the LDA equilibrium 
lattice parameter $a_{\rm th}$, at $a_{\rm exp}$ = 5.16{\AA} we find a smaller
$E_{\rm g}(\Gamma) = 3.92$ eV, {\it i.e.} a reduction of about 0.4 eV.
The behaviour at the $X$ point is different, since the $E_{\rm g}(X)$
calculated at $a_{\rm exp}$ is equal to 3.11 eV,
about 0.1 eV bigger than that computed at $a_{\rm th}$.
Given the different nature of the electronic states 
at the various symmetry points
in the BZ, a non trivial dependence of the eigenvalues upon
the structural parameters is generally expected \cite{note2}.

On the other hand,
the QP energy differences obtained by our method are less dependent on
the precise value of the optical dielectric constant used in the 
model screening (Eq.\ref{eq:6}). For instance, $E_{\rm g}(\Gamma)$
varies only by -3\% when passing from $\epsilon_\infty=3.35$ to
$\epsilon_\infty=3.7$, which represents a 11\% increase of
the optical dielectric constant.

We are not aware of any angle--resolved photoemission or inverse
photoemission experiments on SrO, able to provide direct experimental
information on the band dispersion around the Fermi level. Thus, the
comparison of our results to experimental data is mainly based on the
information issued from reflectivity measurements.
Rao and Kearney \cite{Rao79} conclude for direct gaps at $\Gamma$ ($X$) 
$E_{\rm g}(\Gamma) = 5.9$ eV ($E_{\rm g}(X) = 6.28$ eV), and exciton 
binding energies at high--symmetry points of the BZ around 0.2 - 0.3 eV.
Conversely, more recent optical measurements on well characterized
single crystals of SrO \cite{Kaneko88} found $E_{\rm g}(X) = 5.79$
eV,  lower than  $E_{\rm g}(\Gamma) =  6.08$ eV.
On the basis of the onset of the optical absorption spectra the authors 
also deduce that SrO has an indirect minimum gap, while BaO shows a
direct gap at $X$. 
On the theoretical side, the nature of the fundamental gap of SrO
is still debated: while both LMTO/LDA \cite{Taurian}
and APW/X$_\alpha$ \cite{Hasegawa80} calculations give an indirect gap 
$\Gamma_{15}-X_3$ clearly underestimated (3.8 and 3.9 eV respectively),
Pandey and coworkers \cite{Pandey91} found a more realistic value
$E_{\rm g}(\Gamma) = 7.11$ eV  and $E_{\rm g}(X) = 9.11$ eV
through a Hartree--Fock calculation including
correlation at the second order in perturbation theory.
These results confirm the general findings in other semiconducting and
insulating materials, that DFT--LDA underestimates the fundamental gap,
while the reverse happens in Hartree--Fock calculations, although 
the inclusion of correlations in the latters
improves the agreement with experiments.
Our GW corrected bands at the experimental lattice parameter, yield 
gaps at $X$ and $\Gamma$  equal to $E_{\rm g}(X) = 6.45$ eV and 
$E_{\rm g}(\Gamma) = 7.12$ eV respectively, presenting a better
agreement with experimental data.

\begin{table}[p]
\begin{tabular}{|l|c|c|c|}
\noindent
SrO  & $a$({\AA}) & $B$(Mbar) &$E_{\rm coh}$ (eV) \\
\hline
{\rm Experiment}        & 5.16 & 0.906 & 10.45 \\
\hline
{\rm HF \cite{Zupan} }         & 5.25    & 1.06      &         \\
{\rm LDA \cite{Villafiorita}}  & 5.1-5.18& 0.88-0.82 & 10.9-10.5 \\
{\rm LDA \cite{Taurian}}       & 5.22    & 1.07      & 9.59 \\
\hline
{\rm Present work}             & 5.07    & 1.04      & 11.9 \\
\end{tabular}
\caption{ Structural properties of Cubic SrO. A comparison is
made with Hartree--Fock (Ref.\protect\onlinecite{Zupan}) and LDA calculations
(Refs.\protect\onlinecite{Villafiorita,Taurian}).  }
\label{tab:2a}
\end{table}

\begin{table}[h]
\begin{tabular}{|l|c|c|c|c|}
\noindent
%{\rm Cubic SrO}         &{$\Gamma-\Gamma$} &{X-X} &{L-L} &{Minimal Gap}
 SrO  & $E_{\rm g}(\Gamma)$ & $E_{\rm g}(X)$ & $E_{\rm g}(L)$ 
& {Minimal Gap}
\\
\hline
Experiment                  & & & & \\
Ref.\onlinecite{Rao79}      & 5.896& 6.28 &     & 5.9 ($\Gamma-\Gamma$)\\
Ref.\onlinecite{Kaneko88}   & 6.08 & 5.79 &     & \\
\hline
LDA (Ref.\onlinecite{Taurian})       & 4.26 & & & 3.8($\Gamma-X$)\\
$X_\alpha$ (Ref.\onlinecite{Hasegawa80})  &5.10&4.03&7.3&3.9($\Gamma-X$)\\
HF+corr. (Ref.\onlinecite{Pandey91}) &7.11&9.11&12.36&8.54($\Gamma-\Gamma$)\\
\hline
Present work      & & & & \\
LDA$^\dag$        & 4.29 & 3.03 & 7.79 & 3.00 ($\Gamma-X$)\\
LDA$^\ddag$       & 3.92 & 3.11 & 7.47 & 3.04 ($\Gamma-X$)\\
LDA+GW$^\dag$     & 7.54 & 6.39 & 11.17 & 6.37 ($\Gamma-X$)\\
LDA+GW$^\ddag$    & 7.12 & 6.45 & 10.81 & 6.39 ($\Gamma-X$)\\
\end{tabular}
\caption{Gap of Cubic SrO (in eV), at the high--symmetry point of the
Brillouin zone. The gaps are calculated either at the theoretical
lattice parameter ($^\dag$) or at the experimental one ($^\ddag$).}
\label{tab:2b}
\end{table}

\section{{\bf Strontium Titanate}}

\begin{figure}[h]
% \special{psfile=SrTiO3_bands.ps hoffset=-30 voffset=40 
%                                 hscale=42 vscale=45 angle=270}
% \vspace*{7.3cm}
 \includegraphics{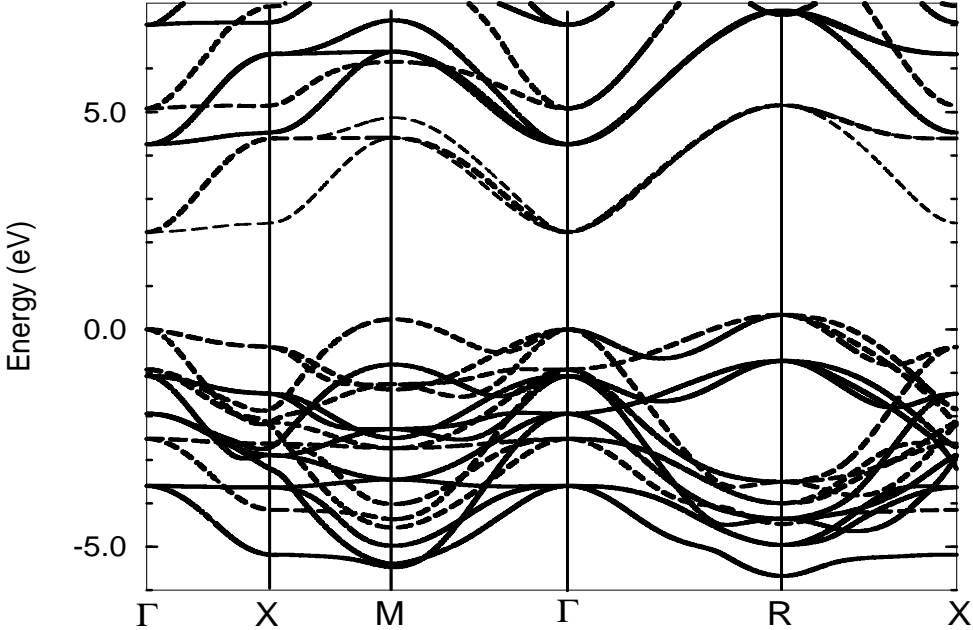}
 \vspace*{9.3cm}
 \caption{ Upper part of the valence bands and lower part of the
conduction bands of cubic SrTiO$_3$, at the theoretical equilibrium
lattice parameter. Dashed lines: LDA bands.  Solid line: GW bands. }
 \label{fig:srtio3}
 \end{figure}

The computed ground--state properties of the cubic phase of
SrTiO$_3$, such as the equilibrium 
lattice parameter, the bulk modulus and the cohesive energy, are 
showed in Table \ref{tab:3a}. A good agreement with previous LDA
calculations \cite{Kimura,King} is found. We stress that in all these
calculations the selfconsistent electron density includes the semicore states 
(Sr $4s$, Sr $4p$, Ti $3s$ and Ti $3p$). With respect to experimental
data, a slight underestimate (-1.3\%) of the lattice parameter is
obtained, while the computed cohesive energy is $\simeq 20\%$ 
larger than that measured, as it is usual in the LDA. As far as the GW
calculation is concerned, we adopt the experimental value 
$\epsilon_\infty = 5.82$ of the optical dielectric constant of the cubic
phase of SrTiO$_3$ \cite{Dore}. Both LDA and GW bands 
(Fig.\ref{fig:srtio3}) are computed at the LDA
theoretical equilibrium lattice parameter $a_{\rm th}$.

Firstly, we discuss the positions of the deep lying states not shown in
Fig.\ref{fig:srtio3}.
A deep, almost non dispersive, energy band is found at 
$\simeq -54$eV below the top of the valence band (VBM)
in the LDA calculations, which originates from Ti $3s$ states.
The inclusion of QP corrections in our GW scheme
pushes it downward, at -58.9 eV below the VBM.
This is in very good agreement with XPS measurements \cite{Kowal77}, 
which show a peak around -59 eV.

Analogously, in the LDA calculation, the narrow bands
originating mainly from Ti $3p$ states  are located
at -30.5 eV below VBM, very close to a band with dominant Sr $4s$
character, at about -29.5 eV. GW corrections push them at -33.7 eV 
and -32.7 eV, respectively. 
The XPS spectra show a 2 eV wide asymmetric peak 
centered at $\simeq -34.5$ eV \cite{Kowal77}.
As found for the Ti $3s$ band,
the inclusion of quasiparticle effects greatly improves the
agreement with the experimental data.

The very broad structure in the XPS data \cite{Kowal77} between 
-20 eV and -14 eV originating form O $2s$ and Sr $4p$ states 
is fully consistent with the position
of our GW bands, while the LDA bands are higher in energy
by about 3 eV on average.

The valence band portion shown in Fig.\ref{fig:srtio3}
has a dominant O $2p$ character with
non negligible contributions from Ti $3d$ states, mainly in its lower
part. It is less sensitive to GW corrections,
which almost rigidly shift the bands towards lower energies. As a
consequence, the LDA and GW bandwidths along the $\Delta$ line
(4.15 eV and 4.11 eV, respectively) are both in very good agreement with
the value of 4.2 eV measured in angle--resolved photoemission \cite{Brookes}.
Similarly, the total bandwidth 
$R_{1,v}-R_{12,v}$ (see also Fig.\ref{fig:srtio3})
turns out to be 4.8 eV in the LDA and 4.9 eV with quasiparticle corrections.

A major difference between the LDA and GW results regards the direct
band gaps, which roughly differ by about 3 eV. The onset for optical
absorption was measured at 3.2 eV by Cardona \cite{Cardona} and
at 3.34 eV by Blazey \cite{Blazey}, and interpreted as a $\Gamma_{15,v}
\rightarrow \Gamma_{25^\prime,c}$ transition. The strong
absorption peaks at $\simeq 4-5$ eV were attributed to 
vertical transitions at $X$ and $M$, respectively. 
On the other hand,
the fundamental gap derived from combined photoemission and 
inverse photoemission spectra, which is free from excitonic effects, 
was estimated to be $(3.3\pm 0.5)$ eV \cite{Tezuka94}.
With respect to these experimental values,
the LDA $E_{\rm g}(\Gamma)$ is underestimated by almost 1 eV, 
while the GW one is overestimated by about 2 eV (see Table \ref{tab:3a})
In this respect, 
the inclusion of self-energy corrections at a perturbative level
does not improve the LDA. 
This issue is discussed in the next
section, in relation to the validity of our model screening for
transition metal oxides such as SrTiO$_3$.

\begin{table}[h]
\begin{tabular}{|l|c|c|c|}
\noindent
 SrTiO$_3$ & $a$({\AA}) & $B$(Mbar) & $E_{\rm coh}$(eV) \\
\hline
Experiment              & 3.903 & 1.83 & 31.7 \\
\hline
Ref.\onlinecite{Kimura} & 3.870 & 1.94 & \\
Ref.\onlinecite{King}   & 3.864 & 1.99 & \\
This work               & 3.850 & 2.03 & 37.88 \\
\end{tabular}
\caption{ Structural properties of cubic SrTiO$_3$. Both
Refs.\protect\onlinecite{Kimura,King} use the LDA with ultra--soft
pseudopotentials. The experimental values of $a$ and $E_{\rm coh}$
are quoted in Ref.\protect\onlinecite{Weyrich} and $B$ is taken
from Ref.\protect\onlinecite{Fischer}. }
\label{tab:3a}
\end{table}

\begin{table}[b]
\begin{tabular}{|l|c|c|c|c|c|}
\noindent
 SrTiO$_3$ & $E_{\rm g}(\Gamma)$ & $E_{\rm g}(X)$ 
& $E_{\rm g}(M)$ & $E_{\rm g}(R)$ & Minimal Gap \\
\hline
LDA \cite{Kimura}    & 2.16 &      &     &      & 1.79 ($R\!-\!\Gamma$) \\
LDA \cite{Xu}        & 3.77 & 4.1  &     & 5.2  & 2.88 ($R\!-\!\Gamma$) \\
\hline
{\rm This work}         & & & & & \\
LDA                  & 2.24 & 2.85 & 4.17 & 4.82 & 1.90 ($R\!-\!\Gamma$)\\
LDA+GW               & 5.42 & 6.10 & 7.29 & 8.15 & 5.07 ($R\!-\!\Gamma$)\\
\end{tabular}
\caption{Computed $E_{\rm g}$ for direct transitions 
from the VB to the CB of cubic SrTiO$_3$, in eV. 
The minimal gap is also given. 
}
\label{tab:3b}
\end{table}

% \begin{figure}[p]

\section{Discussion}

The agreement of our numerical results with the available experimental data
is rather inhomogeneous:
while for MgO and SrO a reasonable agreement is found, for SrTiO$_3$
the QP bands around the Fermi level seem to be worse than the LDA ones. 
It is the aim of this section to discuss our results with respect to two
points:
On the one hand, the quality of the ground--state calculation from which the
perturbative GW calculation starts. On the other hand, the
reliability of our model dielectric function used in the calculation of 
the QP spectra of the oxides considered here.

We have found that the 
values of the direct gaps are very sensitive to the  value of the
lattice parameter. A good description of the ground state structural
properties is thus a prerequisite to obtain QP spectra
that can be compared to experimental data reliably.
This is especially the case of MgO and SrO.
In Table \ref{tab:degap} the relative variations $\Delta_{\rm g}({\vec k})$ 
of the direct gaps, both in the LDA and in the GW
approximation, are given  as a function of the lattice parameter.
One can note that: 
{\it (i)} The largest variations of the direct gaps
$\Delta_{\rm g}({\vec k})$ are nearly always obtained at the LDA level.
{\it (ii)} $\Delta_{\rm g}({\vec k})$ depends on the actual ${\vec k}$ 
point, consistently with the character of the wavefunction. The largest
$\Delta_{\rm g}({\vec k})$ are generally found for the strongest
bonding--antibonding combinations across the gap.
{\it (iii)} Increasing $\Delta_{\rm g}(\Gamma)$ are found
for SrTiO$_3$, SrO and MgO, in ascending order.

One should thus be careful when discussing the quality of the self-energy
corrections to the electronic structure, since it may be strongly
dependent on variables, such as the lattice parameter, which are 
{\it external} to the theory itself. Our data for MgO, obtained at the
 experimental lattice constants $a_{\rm exp}$, 
compare well with other more sophisticated, ab--initio GW results 
also using $a_{\rm exp}$ \cite{Schoenberg,Shirley},
as far as the direct gaps are concerned (see Table \ref{tab:1b}). 
Both for MgO and SrO, a fair accordance with experimental data 
is found when adopting $a_{\rm exp}$.
We thus conclude that, still at the level of the ground state structural
properties, one should obtain a good agreement with experimental data
(for example, by using generalized gradient corrections to the LDA) 
before starting the calculation of the quasiparticle energies.

Care must also be taken when comparing the theoretical gaps and 
those derived from optical measurements, because of 
the existence of excitonic effects in the latters, 
not accounted for in the present theory. For MgO,
these effects should be rather small, as suggested experimentally
by Roessler and Walker \cite{Roessler} and calculated
by Benedict and coworkers \cite{Benedict98}.  For SrO and SrTiO$_3$
there is still a lack of {\it ab initio} calculations of one-- and
two--particle excitations to address fully this issue.

Another important point is the quality of the
quasiparticle corrections, in relation to the model dielectric function
$\epsilon(q,\tilde\rho)$ used in the GW calculation. 
As it is evidenced by Eq.\ref{eq:6}, the screening depends on the value of
the optical dielectric constant $\epsilon_\infty$ of the oxide. 
As in the previous discussions regarding the lattice parameter, 
one can use either the experimental optical dielectric constant 
available in the literature, or compute it in the framework of the
self--consistent theory adopted, such as the DFT. In a recent paper, Shirley
\cite{Shirley} obtained for MgO a theoretical optical dielectric
constant, at the {\it experimental} lattice parameter, 
$\epsilon_\infty^{\rm (th)}=3.03$ in the LDA, to be compared 
with the measured $\epsilon_\infty=2.95$ \cite{Whited}.
Extrapolating our results -- obtained with $\epsilon_\infty=2.95$ and
$\epsilon_\infty=3.50$ (see Sec.IV) -- we can conclude that our 
quasiparticle energies would vary by less than 0.1 eV when passing from
$\epsilon_\infty=2.95$ to $\epsilon_\infty^{\rm (th)}=3.03$.
However, a more satisfactory theoretical framework would be to compute 
$\epsilon_\infty$ from first principles, in conjunction with the use of an 
approximation for the exchange--correlation energy, capable to reproduce
the experimental lattice constant. Although in the present case the 
results would not be much affected, this would open the way to the
calculation of quasiparticle excitations of systems for which no
experimental measurements of the optical dielectric constant are
available.  
On the other hand, the use of model dielectic functions
might open the possibility to treat complex systems consisting of
many inequivalent atoms, such as low--symmetry crystals,
surfaces and heterostructures.

Considering the numerical results obtained, and the previous
discussion, we note that the model dielectric function
works reasonably well for both MgO and
SrO, which are compounds with fundamental gaps much larger than 
the semiconductors ZnSe, GaN, and SiC previously 
studied with the same method \cite{Palummo95,Wenzien1}.  
We thus conclude that the
validity of the screening model (Eq.\ref{eq:6}) is neither a function of the
gap value nor of the ionocovalent character of the bonding in 
the crystal under study.
On the other hand, it fails when applied
to SrTiO$_3$, a transition metal oxide characterized by a moderate
optical gap (slightly larger than 3 eV according 
to Refs.\onlinecite{Brookes,Blazey}). 

In order to clarify the reasons for such a failure, we computed the
bare exchange contribution to the QP gap at the $\Gamma$ point, for 
both MgO (10.46 eV) and SrTiO$_3$ (12.28 eV), in first--order pertubation
theory. This difference stems from the different character of the states
at the bottom of the conduction bands, which have a Ti 3$d$ character in
the case of SrTiO$_3$, and are much flatter than those of MgO, having
a dominant Mg 3$s$ character. This is fully consistent with the stronger
localization of the Ti 3$d$ states, which could make our treatment of
local--field effects (see Eqs.\ref{eq:6} and \ref{eq:7}) rather
inappropriate. In fact, the choice of the effective density 
$\tilde\rho_{n \vec k}$ (Eq.\ref{eq:8}) takes into account the different
localization of the states only through their superposition with the
electron density $\rho(\vec r)$. Moreover, its definition contains an
ambiguity, since its precise value depends on the core--valence
partitioning of $\rho(\vec r)$.
It is interesting to consider the extreme limit in which 
$\epsilon(q,\tilde\rho)$ is no longer $q$ dependent and is
simply replaced by the optical dielectric constant
$\epsilon_\infty$. 
In this limit of screening, the
only contribution to the GW correction to the DFT--LDA gap  
comes from the screened--exchange term, since the Coulomb hole 
gives a rigid shift equal for all the bands. 
Interestingly, this correction provides gap values equal to 4.3 eV for
MgO and 2.7 eV for SrTiO$_3$, in much better agreement with experiments
in the latter case than when using 
$\epsilon(q,\tilde\rho_{n \vec k})$ given by 
Eq.\ref{eq:6}. This points out the need for a better inclusion of local
field effects in the model dielectric function for transition metal
oxides, such as SrTiO$_3$.
Another source of failure relatively to SrTiO$_3$ could be also the use
of a perturbative GW scheme: a possible solution to this issue would be
the use of an efficient self-consistent GW approach which has shown to
give good results even in systems in which $d$ orbitals play a fundamental
role \cite{Massidda}.

\begin{table}[p]
\begin{tabular}{|l|c|c|c|}
\noindent
\hspace*{3cm} & MgO & SrO & SrTiO$_3$ \\
\hline
$(a_{\rm th}-a_{\rm exp})/a_{\rm exp} (\%)$ & -2.0 &  -1.7 & -1.3 \\
\hline
$\Delta_{\rm g}^{\rm LDA}(\Gamma)$ & -6.4 & -5.4 & -1.7 \\
$\Delta_{\rm g}^{\rm GW}(\Gamma) $ & -4.1 & -3.6 &  \\
\hline
$ \Delta_{\rm g}^{\rm LDA}(X)$ & -0.7 & 1.5 & -2.2 \\
$ \Delta_{\rm g}^{\rm GW}(X) $ & -0.7 & 0.6 & \\
\hline
$ \Delta_{\rm g}^{\rm LDA}(L)$ & -2.7 & -2.5 & -4.4 \\
$ \Delta_{\rm g}^{\rm GW}(L) $ & -2.1 & -2.0 & \\
\end{tabular}
\caption{Dependence of the direct gaps $E_{\rm g}(\vec k)$
of MgO, SrO and SrTiO$_3$ at high-symmetry points $\vec k$
of the Brillouin zone as a function of the relative variations
of the lattice parameter $(a_{\rm th}-a_{\rm exp})/a_{\rm exp}$.
The function $ \Delta_{\rm g}$ is defined as
${E_{\rm g}(a_{\rm th})- E_{\rm g}(a_{\rm exp}) \over E_{\rm g}(a_{\rm exp})}
{a_{\rm exp} \over a_{\rm th} - a_{\rm exp}} $.
The values are computed either by using the LDA, 
or by including quasiparticle corrections within the GW approximation.}
\label{tab:degap}
\end{table}

\section{Conclusions}

We have calculated the gound state properties of cubic oxides MgO, SrO
and SrTiO$_3$ within the DFT-LDA, and applied a perturbative GW method, 
based on a dielectric model which includes approximately 
local field and dynamical effects, to determine their
quasiparticle energies.  We have tested the dependence of the 
calculated spectra on the parameters entering the calculation,
such as the optical dielectric constant $\epsilon_\infty$ 
and the lattice parameter $a$.
In particular, we have shown that quasiparticle energies
in alkaline-earth oxides are sensitive functions of the lattice parameter, 
even at the LDA level. As a consequence,
the use of either the experimental or the theoretical values for $a$
can influence the quality of the results whenever $a_{\rm th}$ differs 
from $a_{\rm exp}$.

Through a careful comparison with the available experimental data and
previous ab--initio GW calculations, we have showed that the simplified 
GW method works reasonably well for systems, such as MgO and SrO, 
in which the electronic states
of the valence band and the bottom of the conduction band
mainly consist of $s$-- and $p$--like states.
Moreover, in the three oxides, 
GW corrections have been found to be crucial to account for the
energies of semicore electron states, such as O $2s$, Ti $3s$, Ti $3p$,
Sr $4s$ and Sr $4p$, which agree well with the peak
positions recorded in photoemission spectra.
When localized $d$--like states contribute to bands around the Fermi
level, our method produces an overestimation of the self-energy corrections.
This is the case of SrTiO$_3$, for which it is not clear 
at this stage whether a higher order perturbative approach, in which
the LDA wavefunctions are renormalized, is needed, or whether
this failure calls for a more refined treatment of local--field effects. 
%is a matter of our approximate treatment of local--field effects. 

\vspace*{0.5cm} \noindent
We thank 
Lucia Reining for fruitful discussions and careful reading of the manuscript, 
Tristan Albaret for his help in generating the pseudopotentials, 
and Eric Shirley for kindly providing us unpublished results on 
self-energy corrections for MgO.
Computing resources were granted by the IDRIS/CNRS computational centre in
Orsay, under project 980864.
G.C. acknowledges support from the Commission of the European
Communities (framework $IV$) programme under the Networks scheme.

%\end{multicols}


\begin{thebibliography}{}
\bibitem{note_dec}
We used both a Cray C90 computer and a Digital $\alpha$-500 
workstation, with 512 MB of memory.
\bibitem{note2}
In SrO, the bottom of the conduction band $X_{3,c}$ has mainly 
Sr $4d$ character, while at $\Gamma_{1,c}$ the Sr $5s$ character prevails.
In any case, a noticeable degree of second--neighbour hybridization
is present.
\end{thebibliography}
\end{document}